\documentclass[reprint, amsmath,amssymb, aps, prb, superscriptaddress, nofootinbib, nobibnotes]{revtex4-1}

\usepackage{graphicx}
\usepackage{dcolumn}
\usepackage{bm}
\usepackage{scrextend}
\usepackage{hyperref}
\usepackage{tabularx}
\usepackage{amsmath}
\usepackage{relsize}
\newcommand{\one}{{\ensuremath{\pmb{k}}}}
\newcommand{\two}{{\ensuremath{\pmb{q}}}}
\usepackage{filecontents}
\usepackage{color}
\usepackage{multirow}

\begin{document}


\title{Temperature effects on the electronic band structure of PbTe from first principles}

\author{Jos\'e D. Querales-Flores}
\email{jose.querales@tyndall.ie}
\affiliation{Tyndall National Institute, Lee Maltings, Dyke Parade, Cork T12 R5CP, Ireland}
\author{Jiang Cao}
\affiliation{Tyndall National Institute, Lee Maltings, Dyke Parade, Cork T12 R5CP, Ireland}
\affiliation{School of Electronic and Optical Engineering, Nanjing University of Science and Technology, China}
\author{Stephen Fahy}
\affiliation{Tyndall National Institute, Lee Maltings, Dyke Parade, Cork T12 R5CP, Ireland}
\affiliation{Department of Physics, University College Cork, College Road, Cork T12 K8AF, Ireland}
\author{Ivana Savi\'c$^{1,}$}
\email{ivana.savic@tyndall.ie}

\date{\today}

\begin{abstract}
We report a fully {\it ab-initio} calculation of the temperature dependence of the electronic band structure of PbTe. We address two main features relevant for the thermoelectric figure of merit: the temperature variations of the direct gap and the difference in energies of the two topmost valence band maxima located at L and $\Sigma$. We account for the energy shift of the electronic states due to thermal expansion, as well as electron-phonon interaction computed using the non-adiabatic Allen-Heine-Cardona formalism within density functional perturbation theory and the local density approximation. We capture the increase of the direct gap with temperature in very good agreement with experiment. We also predict that the valence band maxima at L and $\Sigma$ become aligned at $\sim 600-700$ K. We find that both thermal expansion and electron-phonon interaction have a considerable effect on these temperature variations. The Fan-Migdal and Debye-Waller terms are of almost equal magnitude but have an opposite sign, and the delicate balance of these terms gives the correct band shifts. The electron-phonon induced renormalization of the direct gap is produced mostly by high-frequency optical phonons, while acoustic phonons are also responsible for the alignment of the valence band maxima at L and $\Sigma$.

\end{abstract}


\maketitle

\section{Introduction}\label{introduction}

Given their ability to convert waste heat into electricity~\cite{review1,Snyder2008,Heeaak9997,kornelius2017}, thermoelectric (TE) materials could, in principle, play an important role in the future development of energy harvesting technologies. However, their practical applications are limited due to poor efficiency, which is extremely challenging to enhance  because of the conflicting requirements for the desired physical properties. A high-performance TE material has to be a good electrical conductor, a poor thermal conductor and, at the same time, possess a large Seebeck coefficient~\cite{review1,Snyder2008}. All these parameters are strongly dependent on the relative energies of the electronic band states relevant for charge transport~\cite{Sofo1994,pei2011}. If the band energy differences are small ($\sim 0.1$~eV), they may be strongly renormalized by temperature~\cite{Gibbs2013,Monserrat2016,Antonius2016}, which in turn may significantly affect thermoelectric transport properties~\cite{Bilc2006}.

PbTe is among the most efficient bulk thermoelectric materials for temperatures between $500$~K and $900$~K~\cite{Snyder2008}. It is a direct narrow-gap semiconductor, and its direct gap, located at the L point, is very sensitive to temperature variations~\cite{Gibbs2013,gibson1952,tauber1966,saakyan1966,Baleva1990}. Interestingly, PbTe exhibits a temperature induced shift of the direct band gap that is opposite to the majority of semiconductors: the gap increases with temperature from 0.19 eV at 30 K to 0.38 eV at 500 K~\cite{Gibbs2013,gibson1952}. The positive temperature coefficient of the direct gap may be beneficial for the thermoelectric performance of PbTe~\cite{Rowe,Gibbs2013,pei2011}, since a larger gap suppresses bipolar effects caused by intrinsic carrier activation at higher temperatures~\cite{Gibbs2013}. The temperature variation of the direct gap also modifies the effective masses and thermoelectric transport coefficients, which can be seen from the non-parabolic two-band Kane model~\cite{Kane1957,RAVIC1971}. It is thus essential to account for the temperature dependence of the direct gap and effective masses when modelling electronic and thermoelectric transport in PbTe and other direct narrow-gap semiconductors~\cite{Bilc2006}.

Furthermore, PbTe has a complex valence band (VB) structure~\cite{Parker2013,kohn1973,Albanesi2000,zunger1997,rabii}, with two energetically close maxima whose energy difference also depends on temperature. The top VB maxima are located at the L point, while additional heavier pockets are situated along the $\Sigma$ line, with the maximum at $\Sigma = \dfrac{2\pi}{a}(3/8,3/8,0)$, where $a$ is the lattice constant. The secondary VB maximum (VBM) at $\Sigma$ is considered to be $\sim 0.1-0.17$ eV below that of the VBM at L at low temperatures ($\sim 4$ K)~\cite{allgaier1961,allgaier1966,riedl1965,sitter1977,Lusakowski2011,singh2010}. It has been argued in the literature that the energy difference between these two VBM decreases due to temperature, and they become aligned at a temperature between $400$~K and $700$~K~\cite{Gibbs2013,kaviany2012,sitter1977}. This feature of the electronic band structure of PbTe is of great interest for improving its thermoelectric performance. It has been suggested that such enhanced band degeneracy, induced by temperature or alloy composition, may yield higher Seebeck coefficient without significantly reducing electronic conductivity~\cite{pei2011,Gibbs2013,zhu2014}. To accurately account for the influence of this valence band alignment (or ``band convergence''~\cite{pei2011}) on the thermoelectric performance of $p$-type PbTe, it is necessary to reliably determine the temperature variation of the energy difference between the VBM at L and $\Sigma$.  

The first-principles theoretical framework for calculating temperature dependent electronic band structures based on the Allen-Heine-Cardona (AHC) formalism~\cite{allen1976,allen1981,allen1983,allen1981} and density functional perturbation theory (DFPT)~\cite{abinit2,abinit3,Baroni2001} has been developed recently~\cite{ponce2015jcp,ponce2014}. The temperature dependence of electronic energies originates from thermal expansion and electron-phonon interaction (EPI)~\cite{allen1981,keffer1968,tsang1971,cohen1975}. Allen, Heine and Cardona~\cite{allen1976,allen1981,allen1983,allen1981} developed the theoretical approach that accounts for the renormalization of electronic bands due to EPI, and includes the second-order contributions with respect to atomic displacement, known as the Fan-Migdal and Debye-Waller terms. They showed that EPI can induce renormalization of the band structure comparable to those induced by electron correlations~\cite{allen1976,allen1981,allen1983,allen1981}. The AHC formalism has recently been recast in a form suitable for the first-principles calculations of the EPI contribution using DFPT~\cite{ponce2015jcp,ponce2014}.

In this work, the temperature renormalization of the electronic structure of PbTe due to electron-phonon interaction and thermal lattice expansion is studied from first-principles. We calculate the temperature dependence of the direct gap and the energy difference between the two topmost valence band maxima at L and $\Sigma$. The electron-phonon contribution is computed using the non-adiabatic AHC approach and DFPT combined with the local density approximation, while the thermal expansion contribution is obtained by calculating the electronic band structure of the thermally expanded lattice using density functional theory. We obtain a positive temperature coefficient for the direct gap, $\frac{dE_{g}}{dT}$, that agrees well with experimental results~\cite{Gibbs2013,gibson1952,tauber1966,saakyan1966,Baleva1990}. We predict that the temperature at which the valence band maxima at L and $\Sigma$ ``converge'' is $\sim 600\--700$ K. We show that both thermal expansion and electron-phonon interaction give sizeable contributions to these temperature changes. The sign of the temperature variations of the direct gap and the energy difference between the valence band maxima at L and $\Sigma$ originate from the Debye-Waller and the Fan-Migdal contributions to EPI, respectively, together with thermal expansion. The dominant contribution to the electron-phonon renormalization of the direct gap stems from high-frequency optical phonons, while acoustic phonons also contribute to the ``convergence'' of the valence band maxima at L and $\Sigma$.

\section{Method and computational details}\label{calculation}

\subsection{Ground-state calculations}

We obtain the electronic band structure of PbTe at $0$~K using density functional theory (DFT) and the local density approximation (LDA)~\cite{Caperley1980,Perdew1981} implemented in the {\sc ABINIT} code~\cite{abinit1,Gonze2016}. We use Hartwigsen-Goedecker-Hutter norm-conserving pseudopotentials~\cite{Hartwigsen1998} with the $6s^2 6p^2$ states of Pb and $5s^2 5p^4$ states of Te explicitly included in the valence states. We use the cutoff energy of $45$~Ha, and a 12$\times$12$\times$12 Monkhorst-Pack $\one$-point grid. The spin-orbit interaction (SOI) at the LDA level of theory underestimates the band gap to such a degree that the conduction and valence bands invert and mix heavily near the L point, producing a ``negative'' band gap~\cite{ronan2018}.  In contrast, excluding SOI in the LDA calculations for PbTe leads to the correct character of the conduction and valence band states near the direct gap at L. In this work, we use both the LDA including and excluding SOI to calculate the temperature dependence of the electronic bands of PbTe.

\subsection{Temperature renormalization of electronic bands}

In the finite temperature regime, the temperature ($T$) dependence of a single particle electronic energy is given as $E_{n\one}(T) = \varepsilon_{n\one} + \Delta E_{n\one}(T)$, where $n\one$ is the state index and $\varepsilon_{n\one}$ is the energy in the case where all the atoms are kept frozen in their equilibrium positions at $0$~K. The temperature variation of the electronic energy, $\Delta E_{n\one}(T)$, includes two contributions\cite{allen1981,cohen1975}:

\begin{equation}\label{eq}
\Delta E_{n\one}(T) = \left(\frac{\partial \varepsilon_{n\one}}{\partial T}\right)_{P} = \left(\frac{\partial \varepsilon_{n\one}}{\partial \ln V}\right)_{T} \beta + \left(\frac{\partial \varepsilon_{n\one}}{\partial T}\right)_{V},
\end{equation}

\noindent where the first term represents the energy renormalization due to lattice thermal expansion i.e. the thermally induced change in volume at constant temperature ($\beta$ is the volumetric thermal expansion coefficient). The second term is the energy renormalization due to phonon populations i.e. the vibration of atomic nuclei at constant volume. The effect of electron-phonon interaction at constant volume on the temperature induced energy shifts is usually the dominant term in Eq.~\eqref{eq}, and is the most difficult term to compute from first principles~\cite{ponceprb2014}. We calculate the renormalization of the electronic structure of PbTe due to thermal expansion and electron-phonon interaction as described in the following subsections.

\subsection{Thermal lattice expansion}

 We calculate the effect of thermal expansion on the electronic band structure of PbTe by varying the lattice constant that accounts for thermal expansion of the lattice, and computing the corresponding electronic structures using DFT. We obtain temperature dependent lattice constant including zero-point renormalization (ZPR) as~\cite{pavone1994}:
 
\begin{equation} 
a(T) = a_{0} + \frac{1}{3N_{\two}B}\sum_{\two \lambda}{\hbar \omega_{\two \lambda}\gamma_{\two\lambda}\left(n_{\two \lambda}(T)+\frac{1}{2}\right)}.
\end{equation}
 
\noindent Here $a_{0}$ is the lattice constant calculated using DFT-LDA, $N_{\two}$ is the total number of sampled $\two$-points, $B$ is the bulk modulus, $\omega_{\two \lambda}$ is the frequency of the phonon mode with the wave vector $\two$ and the branch index $\lambda$,  $\gamma_{\two\lambda}$ is the mode Gr\"{u}neisen parameter defined as $\gamma_{\two\lambda}=-d(\log \omega_{\two\lambda})/d(\log V)$~\cite{Srivastava1990} where $V$ is the primitive unit cell volume,  and  $n_{\two \lambda}(T)$ is the Bose-Einstein distribution function for the phonon mode $\two\lambda$ at temperature $T$. We also compute linear thermal expansion coefficient using~\cite{Srivastava1990}
\begin{equation}
\label{th_exp}
\alpha=\frac{1}{3N_{\two}VB}\sum_{\two\lambda}c_{\two\lambda}\gamma_{\two\lambda}, 
\end{equation}
where $c_{\two\lambda}$ is the heat capacity of the phonon mode $\two\lambda$. Phonon frequencies used in the calculation of the lattice constant and linear thermal expansion coefficient of PbTe were computed using harmonic interatomic force constants at 0~K obtained from Hellman-Feynman forces for 128-atom supercells using LDA excluding SOI~\cite{murphy2017}. In Appendix \ref{newappA}, we present the comparison between our calculated thermal lattice expansion, lattice constant and phonon dispersion of PbTe with experiments.

\subsection{Electron-phonon interaction}

The electron-phonon renormalization of the electronic structure of PbTe, including the zero point renormalization, is calculated using the Allen-Heine-Cardona theory~\cite{allen1976,allen1981,allen1983} and its DFPT implementation in the {\sc ABINIT} code~\cite{abinit1,Gonze2016}. The main aspects of the AHC approach are summarized as follows~\cite{ponce2015jcp,Giustino2017}. Electron-phonon interaction is treated perturbatively, and consists of two terms representing the second order Taylor expansion in the nuclear displacement, known as the Fan-Migdal (FAN) and Debye-Waller (DW) self-energy terms~\cite{cannuccia2013,antonius2015,Giustino2017}:   
\begin{widetext}
\begin{equation}\label{selfenergy}
\Sigma_{n\one}^{FAN} (\varepsilon_{n\one}, T) = \sum_{n' \two \lambda} \frac{\vert g_{nn'\one}^{\two \lambda}\vert^{2}}{N_\two}    \times\left[  \frac{n_{\two \lambda} (T) +1 -f_{n' \one + \two}(T)}{\varepsilon_{n\one} - \varepsilon_{n' \one + \two} - \omega_{\two \lambda} + i\delta }  + \frac{n_{\two \lambda} (T) + f_{n' \one + \two}(T)}{\varepsilon_{n\one} - \varepsilon_{n' \one + \two} + \omega_{\two \lambda} + i\delta} \right],
\end{equation}

\begin{equation}\label{DW}
\Sigma_{n\one}^{DW} (\varepsilon_{n\one}, T) = \frac{1}{N_{\two}} \sum_{\two \lambda} \Lambda_{nn'\one}^{\two \lambda \two ' \lambda '} \left[ 2n_{\two \lambda} (T) + 1 \right].
\end{equation}

\end{widetext}

\noindent Here $ f_{n' \one - \two} (T)$ is the electronic Fermi-Dirac distribution for the electronic state $n'\one-\two$ \footnote{In the current {\sc ABINIT} code implementation, the electronic occupations $ f_{n' \one - \two} (T)$ are taken to be equal to one and zero for the valence and conduction bands, respectively.}, and $\delta$ is an infinitesimal positive number that indicates how to integrate over the singularity in the self-energy integral i.e. principal part for the real part of the integral, and Dirac delta function for the imaginary part. The first-order electron-phonon matrix elements $g_{nn'\one}^{\two \lambda}$ in Eq.~\eqref{selfenergy} represent the probability amplitude for an electron to be scattered by phonons, and are given as~\cite{cannuccia2013,Giustino2017}

\begin{equation}
g_{nn'\one}^{\two \lambda} = \langle u_{n'\one + \two} \vert\partial v^{KS}_{\two\lambda} \vert u_{n\one} \rangle_{\text{uc}},
\end{equation}

\noindent where $u_{\one n}$ and $u_{\one n + \two}$ represent the Bloch part of the wavefunctions for the initial and final electronic states, and the subscript ``uc'' indicates that the integral is carried out within one unit cell. $\partial v^{KS}_{\two \lambda}$ is the first order derivative of the Kohn-Sham potential with respect to the atomic displacements induced by the phonon mode $\two\lambda$ with frequency $\omega_{\two \lambda}$, and is given by~\cite{Giustino2017}:

\begin{equation}
\partial v^{KS}_{\two  \lambda} = \sqrt{\frac{\hbar}{2\omega_{\two  \lambda}}} \sum_{\kappa  \alpha} {\sqrt{\frac{1}{M_{\kappa}}}} e_{\kappa \alpha}^{
\lambda} (\two)\partial_{\kappa  \alpha, \two} v^{KS},
\end{equation}

\noindent where $e_{\kappa  \alpha}^{\lambda} $ is the $\alpha$-th Cartesian component of the phonon eigenvector for an atom  $\kappa$ with mass  $M_{\kappa}$. $\partial_{\kappa  \alpha, \two} v^{KS}$ is the lattice periodic part of the perturbed Kohn-Sham potential expanded to first order in the atomic displacement.

The second-order electron-phonon matrix elements $\Lambda_{nn'\one}^{\two \lambda \two ' \lambda '} $ in the Debye-Waller term given by Eq.~\eqref{DW} are very challenging to compute~\cite{ponce2014}. To overcome this problem, one can use the rigid-ion approximation and rewrite the Debye-Waller term as the product of first-order electron-phonon matrix elements~\cite{allen1976}, which can be obtained from DFPT~\cite{ponce2015jcp,ponceprb2014}. The non-rigid-ion contribution is expected to be small in extended systems~\cite{ponceprb2014}, and is typically neglected in the calculation of the Debye-Waller term~\cite{Giustino2017}.

The non-adiabatic AHC approach described above, where phonon frequencies are explicitly accounted for in Eq.~\eqref{selfenergy}, allows us to calculate energy shifts due to zero-point renormalization for polar materials like PbTe~\cite{Giustino2017,ponce2015jcp,ponceprb2014}. In Appendix~\ref{adiabatic}, we present the convergence studies for the ZPR of the direct gap with respect to the $\two$-grid density and the
parameter $\delta$. Our AHC-DFPT calculations yield apparently converged ZPR shifts using a 48$\times$48$\times$48 $\two$-grid and $\delta \rightarrow 0$. We present ZPR values calculated in this manner in the rest of the paper, as well as finite temperature energy shifts obtained using a 48$\times$48$\times$48 $\two$-grid and $\delta=1$~meV. These calculations, however, do not fully capture the long-range longitudinal optical phonon (polaronic) contribution to energy shifts as $\two\rightarrow 0$. We show a detailed analysis of this contribution in Appendix~\ref{polaron}, and find that an incomplete description of this effect in our DFPT calculations introduces an error of $\sim 10$\% for the ZPR shifts of PbTe. This error in the energy shifts due to polaronic effects decreases with temperature down to $\sim 1$\% at $800$~K. We also note that the adiabatic approximation (i.e.~neglecting phonon frequencies in Eq.~\eqref{selfenergy}) with sufficiently large values of $\delta$ ($\sim 0.1$~eV) gives comparable values of the temperature variations for the direct gap and the energy difference between the valence band maxima at L and $\Sigma$ as the non-adiabatic approach, but cannot give converged values for their ZPRs when $\delta\rightarrow0$~\cite{Giustino2017,ponce2015jcp,ponceprb2014}.

We note that the DFPT implementations of the AHC approach are currently limited to harmonic effects on the temperature renormalization of the electronic states~\cite{Giustino2017,ponce2015jcp,ponceprb2014}. In contrast, anharmonic effects are taken into account in molecular dynamics (MD)~\cite{kaviany2012} and frozen-phonon AHC~\cite{Monserrat2016,antonius2015} calculations, but the coarseness of the Brillouin zone sampling could be a great limitation for their convergence. On the other hand, the DFPT and frozen-phonon implementations of the AHC approach give insight into the relative importance of the Fan-Migdal and Debye-Waller contributions to electron-phonon coupling, as well as the relative contribution of different phonon modes, unlike MD calculations.

\section{Results and discussion}\label{results}

\subsection{Electronic structure of PbTe using the local density approximation}\label{elec_structure_OK}

We first discuss the ability of the LDA calculations to accurately describe the electronic band structure of PbTe. The LDA without SOI reproduces the essential features of PbTe's band structure: the direct narrow gap at the L point and the valence band maximum located along the $\Sigma$ line, see the solid black line in Fig.~\ref{bs}. Our previous work has shown that the LDA excluding SOI correctly captures the ordering of the VBM and conduction band minimum (CBM) at L, as well as the ordering of the VBM at L and $\Sigma$, in contrast to the LDA including SOI~\cite{ronan2018}. Our computed direct band gap using the LDA excluding SOI is $0.5$~eV, and overestimates those obtained from experiment ($0.19$~eV at $4.2$~K~\cite{Dalven1974,strehlow}) and previous hybrid HSE03 functional and a quasi-particle self-consistent {\sc GW} ({\sc QSGW}) calculations, see Table \ref{gaps_comparison}. The calculated energy difference between the lowest CBM and the second lowest CBM at L is $0.45$~eV  using the LDA without SOI, and underestimates the values obtained using the LDA including SOI or higher levels of theory (a hybrid HSE03 functional, {\sc QSGW}) of $\sim 1.2$~eV~\cite{ronan2018,hybrid,GW}. The computed energy difference between the valence band maxima at L and $\Sigma$ using the LDA excluding SOI is $\Delta=0.15$ eV. This value agrees very well with the values extracted from magneto-transport~\cite{sitter1977,allgaier1966} and optical absorption~\cite{riedl1965} experiments ranging from 0.1 to 0.17 eV at low temperatures ($\sim 4$~K)~\cite{sitter1977,allgaier1961,allgaier1966,riedl1965}, and those obtained using HSE03 and {\sc QSGW}, see Table \ref{gaps_comparison}.  

On the other hand, the combination of LDA's tendency to underestimate the band gap and the inclusion of SOI results in an inverted band gap in PbTe~\cite{ronan2018}. SOI causes the valence band maximum to be repelled upward, while the conduction band minimum is repelled downward. The resulting band gap is underestimated to such a degree that the topmost valence band and the bottommost conduction band become interchanged and mix heavily near L~\cite{ronan2018,hybrid}. Also, including SOI pushes the $\Sigma$ valley upward, which becomes the topmost valence band maximum forming an indirect band gap with the conduction band minimum at L~\cite{ronan2018}, at odds with experimental observations. As shown in Ref.~\onlinecite{kohn1973}, the top valence and bottom conduction bands at L of PbTe correspond to the representations L$^{6+}$ and L$^{6-}$, respectively, but that order is inverted in the LDA calculations including SOI. In the same paper, the topmost valence band at  $\Sigma$ is denoted by $\Sigma^{5}$. To account for the correct ordering of all these states in our LDA calculations that include SOI, we define the direct gap at L as $E_{g} = E_{\text{L}^{6-}}-E_{\text{L}^{6+}}$ and the energy difference between the L and $\Sigma$ valence band maxima as $\Delta = E_{\text{L}^{6+}} - E_{\Sigma^{5}}$. Using this notation, we obtain a negative direct band gap of $-0.3$~eV and $\Delta=0.18$~eV using the LDA including SOI. Consequently, the band gap is substantially underestimated in the LDA with SOI in comparison to the more accurate hybrid HSE03 functional and {\sc QSGW} calculations (see Table~\ref{gaps_comparison}), while $\Delta$ is described accurately.

\begin{table}[]
\caption{Direct band gap at L ($E_g$) and the energy of the local maximum at $\Sigma$ with respect to the valence band maximum at L ($\Delta$) for PbTe, calculated using the local density approximation (LDA) without and with spin-orbit interaction (SOI), and compared to previous hybrid HSE03 functional and quasi-particle self-consistent {\sc GW} ({\sc QSGW}) calculations and low temperature ($\sim 4$~K) measurements.}
\begin{tabular}{|c|c|c|}
\hline
                  &        $E_{g}$ (eV)           &    $\Delta$ (eV)                 \\ \hline
Experiment & 0.19~~\cite{Dalven1974,strehlow} & 0.1-0.17~~\cite{allgaier1961,allgaier1966,riedl1965,sitter1977} \\ \hline
 LDA without SOI                 &           0.5  &  0.15             \\ 
  LDA with SOI                &                 -0.30 &  0.18                 \\ 
  HSE03 with SOI\cite{ronan2018}                 &    0.23  &  0.16               \\ 
  QSGW with SOI\cite{GW}                &             0.29  &  0.21              \\ \hline
\end{tabular}
\label{gaps_comparison}
\end{table}

\subsection{Electronic structure dependence on thermal lattice expansion}\label{elec_structure}

\begin{figure}[h]
  \begin{center}
  \includegraphics[width=8.6cm]{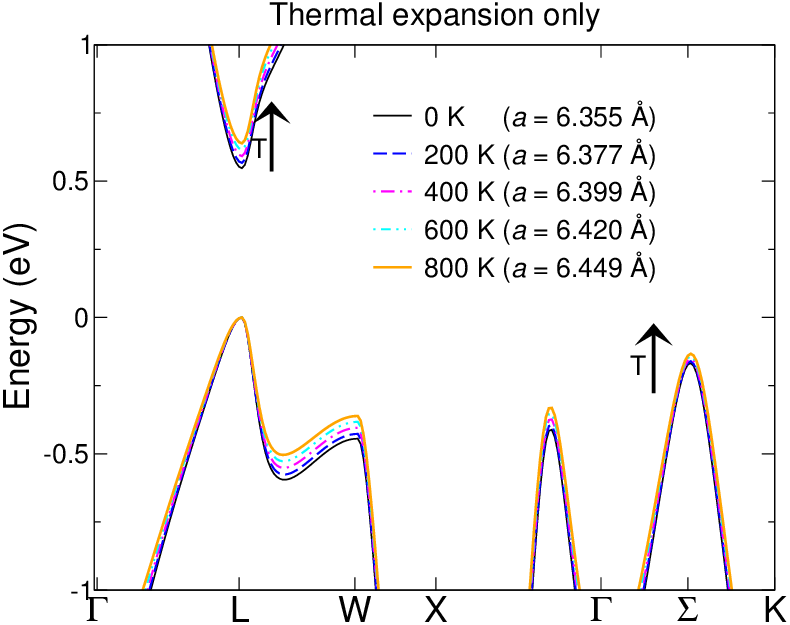}
  \caption[]{Electronic band structure of PbTe for lattice constants at different temperatures, calculated using the local density approximation without spin-orbit interaction, including the effect of thermal expansion and neglecting electron-phonon interaction. The valence band maximum at L is fixed at $0$~eV. Arrows indicate the energy shifts of the states at L and $\Sigma$ as temperature increases.}
  \label{bs}
  \end{center}
  \end{figure}

The temperature dependence of the electronic structure of PbTe due to thermal expansion is shown in Fig.~\ref{bs}, where the VBM at L is fixed at $0$~eV. This and all other figures show our results obtained using the LDA without SOI unless it is explicitly stated that the LDA with SOI is used. We calculate the band structure at the lattice constant values for temperatures ranging from 0 K to 800 K, or equivalently, for the lattice constant expansion up to $1.48$\% with respect to the $0$ K value. Fig.~\ref{bs} clearly shows that thermal expansion increases the direct gap, and reduces the energy difference between the two topmost valence band maxima. We note that thermal expansion renormalizes the direct gap more strongly than the difference in energies between the VBM at L and $\Sigma$.  

\subsection{Temperature variation of the direct gap}\label{eph-direct}

We apply the non-adiabatic AHC theory using DFPT-LDA to compute the zero-point and finite-temperature renormalization of the direct gap of PbTe at L. Details of calculating the ZPR for the direct gap due to electron-phonon interaction (EPI) are given in Appendix~\ref{adiabatic}. We find the direct gap ZPR due to EPI of 19.09 meV and 21.58 meV using the LDA excluding and including SOI, respectively. On the other hand, the ZPR for the direct gap of PbTe due to thermal expansion is 6.58 meV (excluding SOI) and 7.78 meV (including SOI), which translates into the total ZPR of 25.67 meV (excluding SOI) and 29.36 meV (including SOI).

\begin{figure}[h]
  \begin{center}
  \includegraphics[width=8.6cm]{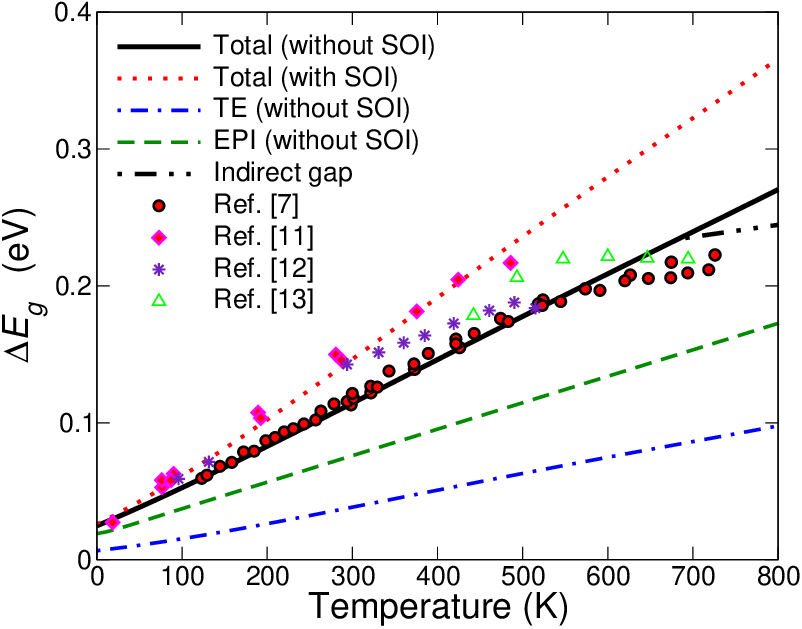}
  \caption[]{Temperature variation of the direct band gap of PbTe with respect to its local density approximation (LDA) value due to both thermal expansion (TE) and electron-phonon interaction (EPI) (solid black and dotted red lines correspond to the LDA excluding and including spin-orbit interaction (SOI), respectively), thermal expansion excluding SOI (dash-dotted blue line) and electron-phonon interaction excluding SOI (dashed green line). Dash-double-dotted black line above $\sim$ 691 K illustrates the crossover from a direct to an indirect gap between the conduction band minimum at L and the valence band maximum at $\Sigma$, calculated excluding SOI. Symbols represent the optical absorption experimental data from Refs.~\onlinecite{Gibbs2013,gibson1952,tauber1966,saakyan1966}.}
  \label{totaldirectgap}
  \end{center}
  \end{figure}

Now we discuss the finite temperature renormalization of the direct gap of PbTe due to both thermal expansion and electron-phonon interaction, and compare it with optical absorption experiments~\cite{Gibbs2013,gibson1952,tauber1966,saakyan1966,Baleva1990}. The total temperature variation of the direct band gap is shown in Fig.~\ref{totaldirectgap}. Due to the inaccurate direct gap values obtained from LDA, we show the temperature dependence of the direct gap with respect to its LDA value, $\Delta E_g=E_{g}(T)-E_{g}(\text{LDA})$, and quantify its temperature derivative $\frac{dE_{g}}{dT}$. Using a linear fit for $E_{g}$ with respect to $T$ in the range of $200\--800$ K, we compute $\frac{dE_{g}}{dT} \approx 3.05 \times 10^{-4}$ eV/K and $\frac{dE_{g}}{dT}\approx 4.35\times 10^{-4}$ eV/K excluding and including SOI, respectively. These values compare very well to the recent experimental value of $\frac{dE_{g}}{dT}\approx 3.2\pm0.1 \times 10^{-4}$ eV/K~\cite{Gibbs2013} obtained from optical absorption data up to 500 K, and the  value of $\frac{dE_{g}}{dT}\approx 4.2 \times 10^{-4}$ eV/K (for $T\leq400$~K) calculated with {\it ab-initio} MD~\cite{kaviany2012}. We note that the experimental gap values continue increasing at a lower rate than $\frac{dE_{g}}{dT}\approx 3.2\pm0.1 \times 10^{-4}$ eV/K for temperatures above $500$~K~\cite{Gibbs2013}. Other values for $\frac{dE_{g}}{dT}$ from optical absorption measurements fall in the range of $\frac{dE_{g}}{dT}\sim 3.0-5.1 \times 10^{-4}$ eV/K~\cite{gibson1952,tauber1966,saakyan1966,Baleva1990}. These studies also report the gap saturation for temperatures above $\sim 500$~K~\cite{gibson1952,tauber1966,saakyan1966}. The likely reason for this non-linear experimental trend is the crossover from a direct to an indirect band gap between the conduction band minimum at L and the valence band maximum at $\Sigma$. This effect in our LDA calculations without SOI is illustrated by dash-double-dotted black line appearing above $\sim$~691 K in Fig.~\ref{totaldirectgap}. Accounting for this crossover, we compute the temperature coefficient for the indirect gap of $\approx 0.88 \times 10^{-4}$~eV/K above $\sim 691$~K (excluding SOI) and $\approx 1.44 \times 10^{-4}$~eV/K above $\sim 623$~K (including SOI).

Our calculated values of $\frac{dE_{g}}{dT}$ using both LDA with and without SOI are within the range of experimental values, and differ from each other by $42$\%. This indicates that accounting for SOI or the correct order of states near the gap does not affect the calculations very much. The reason for this could be that the dominant contribution comes from states that are far away from the gap due to their large density of states. We thus conclude that the accuracy of the electronic band structure does not affect our results more than several tens of percent. Furthermore, all physical trends discussed in this work remain the same regardless of whether SOI is included or excluded in the LDA calculations.

The individual contributions of thermal lattice expansion and electron-phonon interaction to the renormalization of the direct gap of PbTe obtained using the LDA without SOI are also given in Fig.~\ref{totaldirectgap}. We also summarize the individual contributions to the temperature coefficient of $E_{g}$ from thermal expansion and electron-phonon interaction in Table~\ref{slopes}. Both thermal expansion and EPI have a significant effect on $\frac{dE_{g}}{dT}$, and their contributions to $\frac{dE_{g}}{dT}$ are both positive. These findings are in qualitative agreement with those of a recent {\it ab-initio} MD simulation~\cite{Gibbs2013,kaviany2012} and early empirical pseudopotential calculations~\cite{tsang1971,cohen1975}. EPI effects on the direct gap renormalization are stronger than those of thermal expansion up to $800$~K.

\begin{table}[]
\caption{Total and individual contributions to the temperature coefficient of the direct gap ($E_{g}$) and the energy difference between the valence band maxima at L and $\Sigma$ ($\Delta$) from thermal expansion (TE) and electron-phonon interaction (EPI), calculated using the local density approximation (LDA) excluding and including spin-orbit interaction (SOI). These coefficients are obtained using a linear fit for $E_g$ and $\Delta$ with respect to temperature between $200$~K and $800$~K.}
\begin{tabular}{|c|c|c|c|c|}
\hline
\multirow{3}{*}{} & \multicolumn{2}{c|}{\multirow{2}{*}{$\frac{dE_{g}}{dT}$ ($\times 10^{-4}$ eV/K)}} & \multicolumn{2}{c|}{\multirow{2}{*}{$\frac{d\Delta}{dT}$ ($\times 10^{-4}$ eV/K)}} \\
                  & \multicolumn{2}{c|}{}                   & \multicolumn{2}{c|}{}                   \\ \cline{2-5} 
                  & \,\,\, \,\,\, TE \,\,\, \,\,     &  EPI   &  \,\,\, \,\,\, TE \,\,\, \,\,      &  EPI \\ \hline
   LDA without SOI                 &    1.12         &     1.93        &   -0.44         &      -1.73              \\ \hline
    LDA with SOI               &         1.37       &        2.98     &     -0.46     &     -2.43                \\ \hline
\multirow{2}{*}{} & \multicolumn{2}{c|}{\multirow{2}{*}{Total  }} & \multicolumn{2}{c|}{\multirow{2}{*}{Total }} \\
                  & \multicolumn{2}{c|}{}                   & \multicolumn{2}{c|}{}                   \\ \hline
       LDA without SOI & \multicolumn{2}{c|}{3.05}      & \multicolumn{2}{c|}{-2.17}      \\ \hline
LDA with SOI & \multicolumn{2}{c|}{4.35}      & \multicolumn{2}{c|}{-2.89}      \\ \hline   
\end{tabular}
\label{slopes}
\end{table}

\begin{figure}[h]
  \begin{center}
  \includegraphics[width=8.6cm]{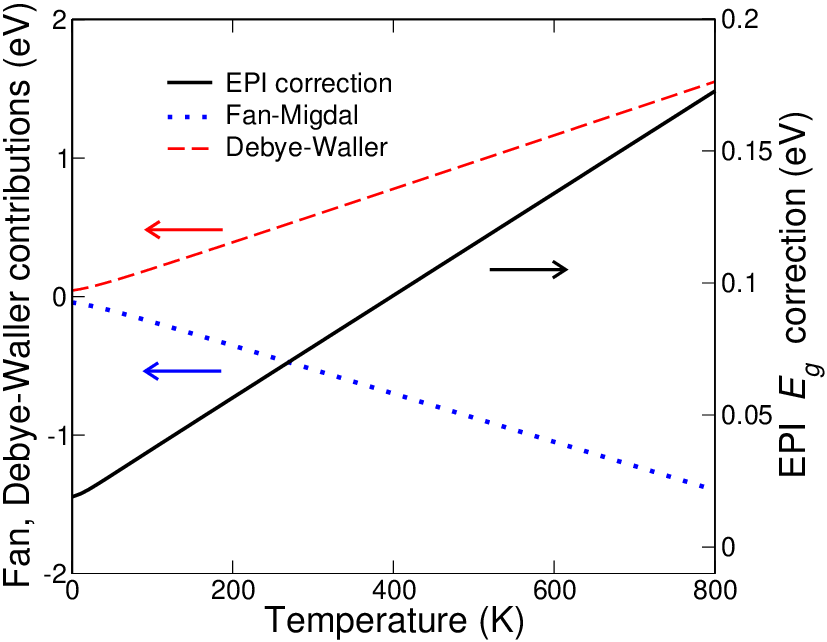}
  \caption[]{Temperature-dependent correction of the direct band gap in PbTe with respect to its local density approximation (LDA) value due to electron-phonon interaction (solid black line), and the Fan-Migdal (dotted blue line) and the Debye-Waller (dashed red line) contributions to electron-phonon interaction. (Note the larger energy scale for the Fan-Migdal and the Debye-Waller terms than for the total correction.) These results are obtained using the LDA excluding spin-orbit interaction.}
  \label{fanvsahc}
  \end{center}
  \end{figure}

Fig.~\ref{fanvsahc} illustrates the effect of electron-phonon interaction on the temperature dependence of the direct gap, together with the Fan-Migdal and Debye-Waller contributions. We find that the Fan-Migdal term reduces the band gap as temperature increases. The Debye-Waller contribution is similar in magnitude to the Fan-Migdal term in PbTe, but it is larger and has the opposite sign. This results in the positive value of $\frac{dE_{g}}{dT}$ i.e. an increasing direct gap with temperature due to EPI in the entire temperature range considered. These trends are in accordance with the conclusions of the detailed theoretical analysis of Ref.~\onlinecite{allen1976} for direct narrow-gap semiconductors. Our results are also consistent with previous empirical pseudopotential calculations in PbTe~\cite{tsang1971,cohen1975,keffer1968,keffer1970} that concluded that the Debye-Waller contribution to the temperature dependence of the direct band gap is significant. Therefore, the Debye-Waller contribution to EPI and thermal expansion both determine the positive sign of the temperature variation for the direct gap of PbTe.
  
\begin{figure}[h]
  \begin{center}
  \includegraphics[width=8.6cm]{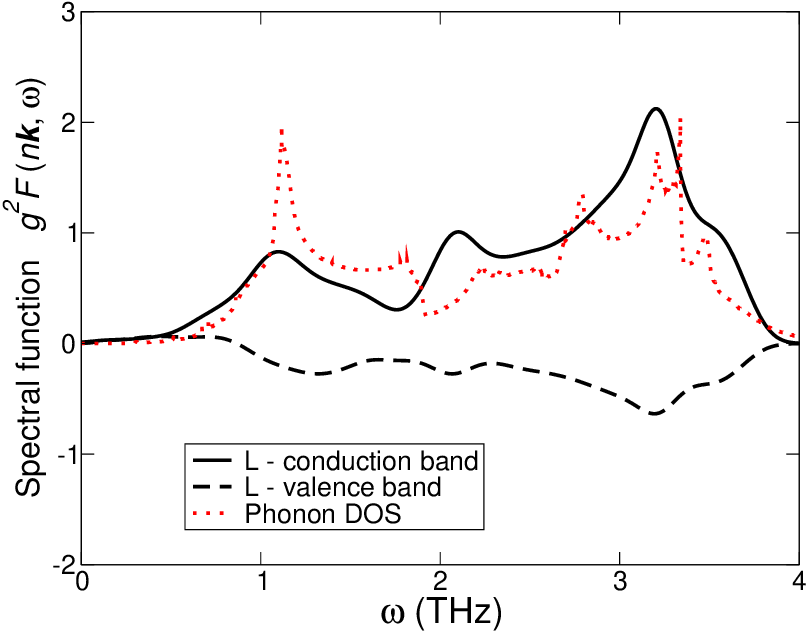}
  \caption[]{Spectral function $g^{2}F(n\one,\omega)$ (see text for explanation) versus phonon frequency for the conduction band minimum (solid black line) and the valence band maximum (dashed black line) at the L point in PbTe. The phonon density of states is also given by dotted red line. These results are computed using the local density approximation excluding spin-orbit interaction.}
  \label{spectraldirect}
  \end{center}
  \end{figure}

We next analyze the frequency-resolved phononic contribution to the electron-phonon renormalization of the VBM and CBM at L, and identify the main contributions. For this, we calculate the spectral function $g^{2}F(n\one, \omega) = \sum_{\two \lambda}\left( \frac{\partial \varepsilon_{n\one }}{\partial n_{\two \lambda}} \right) \delta(\omega - \omega_{\two \lambda})$~\cite{allen1981}, where $\omega_{\two \lambda}$ is the phonon frequency of the mode $\two \lambda$, $\varepsilon_{n\one }$ the electron state energy, and $n_{\two \lambda}$ the phonon population. The spectral function $g^{2}F(n\one, \omega)$ thus represents the phonon density of states weighed by squared electron-phonon matrix elements~\cite{allen1981,garro1996}. We show the spectral functions for the VBM and CBM at the L point in Fig.~\ref{spectraldirect}. The spectral functions are largest for the phonon frequencies between 3 THz and 3.5 THz, which have a dominant effect on the electron-phonon induced renormalization of the direct gap. Comparing the peaks of the spectral functions with the phonon density of states, we find that the largest contribution to the gap changes due to EPI comes from the high-frequency optical phonons.

\subsection{Temperature variation of the topmost valence band maxima at L and $\Sigma$}\label{convergence}

We next use the non-adiabatic AHC theory to calculate the zero-point and finite-temperature renormalization for the energy difference $\Delta$ between the valence band maxima at L and $\Sigma$ in PbTe. We find the ZPR of $\Delta$ due to EPI of -8.72 meV and -9.05 meV using the LDA excluding and including SOI, respectively. The computed ZPR of $\Delta$ due to thermal expansion is -2.37 meV (excluding SOI) and -2.40 meV (including SOI), resulting in the total ZPR of -11.09 meV (excluding SOI) and -11.45 meV (including SOI).

\begin{figure}[t]
  \begin{center}
  \includegraphics[width=8.6cm]{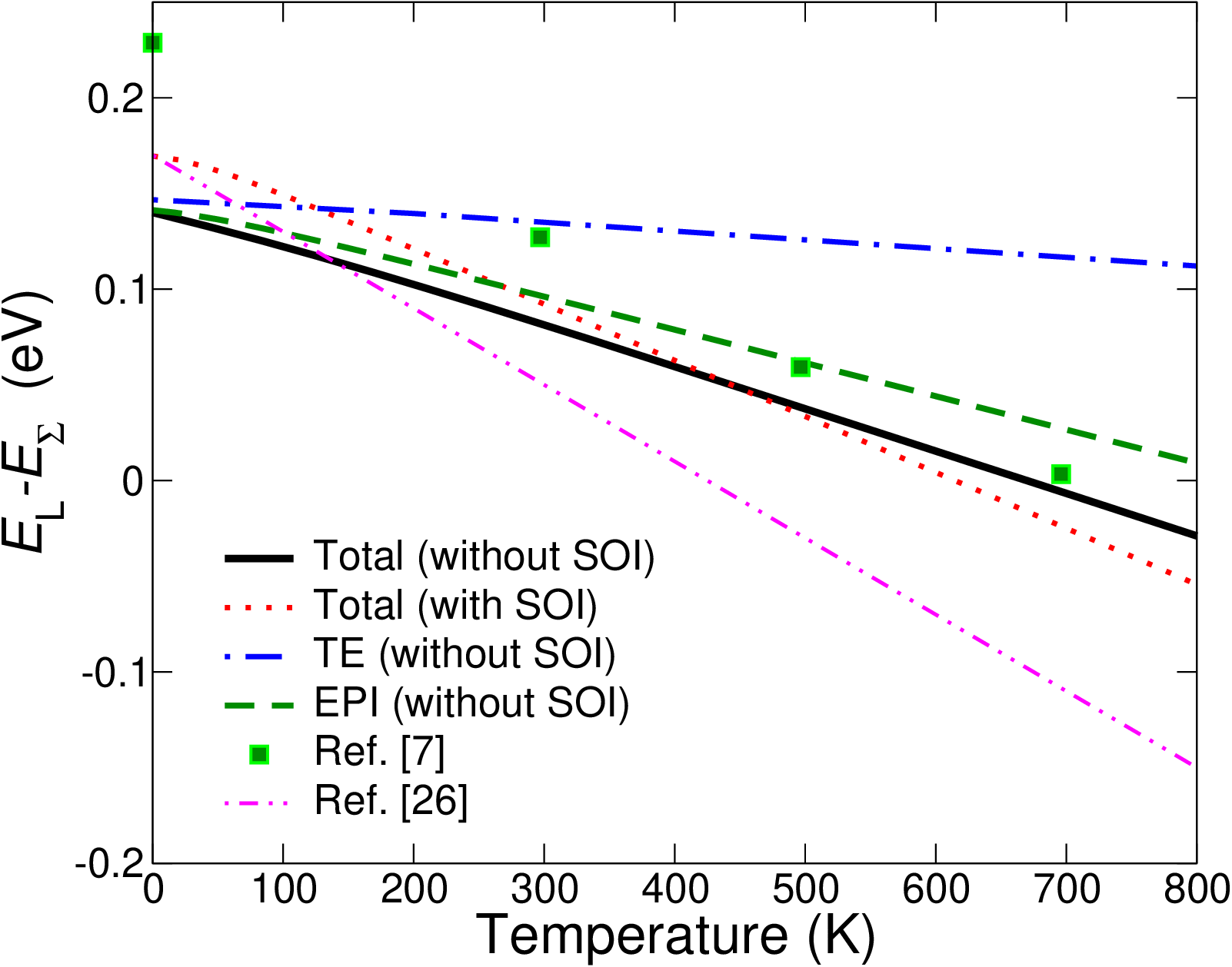}
  \caption[]{Temperature variation of the energy difference between the valence band maxima at L and $\Sigma$ for PbTe due to both thermal expansion (TE) and electron-phonon interaction (EPI) (solid black and dotted red lines correspond to the local density approximation excluding and including spin-orbit interaction (SOI), respectively), thermal expansion excluding SOI (dash-dotted blue line) and electron-phonon interaction excluding SOI (dashed green line). Green squares represent the \textit{ab-initio} molecular dynamics results from Ref.~\onlinecite{Gibbs2013}, while dash-double-dotted magenta line shows the result quoted in Ref.~\onlinecite{sitter1977} and deduced from the temperature saturation of the fundamental gap in optical absorption measurements~\cite{gibson1952,tauber1966}.}
  \label{sigma}
  \end{center}
\end{figure}

The temperature dependence of the energy difference $\Delta$ is illustrated in Fig.~\ref{sigma}. For comparison, we included the corresponding results of an {\it ab-initio} MD calculation~\cite{Gibbs2013} in the same figure. Experimental data for $\Delta(T)$ is scarce, and only infrared reflectivity experiments in Ref.~\onlinecite{riedl1965} reported the $\Delta$ value of $0.08$~eV at 300~K. Most of the literature quotes the temperature coefficient of $\frac{d \Delta}{d T}=-4\times 10^{-4}$ eV/K~\cite{sitter1977}, which was deduced from the temperature saturation of the fundamental gap at $\sim 450$~K observed in optical absorption measurements~\cite{gibson1952,tauber1966}, assuming that this effect indicates the alignment of the VBM at L and $\Sigma$. However, a few recent analyses of the optical and Hall mobility data~\cite{Gibbs2013,Jaworski2013} questioned this result, and concluded that the ``convergence'' of the VBM at L and $\Sigma$ may occur at significantly larger temperatures. We obtain the temperature coefficient of $\frac{d \Delta}{d T} \approx -2.17 \times 10^{-4}$ eV/K and $\frac{d \Delta}{d T} \approx -2.80 \times 10^{-4}$ eV/K using the LDA excluding and including SOI, respectively. These two values differ by $\sim 30$\%, which again confirms that accounting for SOI or the correct order of the states near the gap does not affect the calculations much. We find that the valence band maxima at L and $\Sigma$ ``converge'' at $\sim 691$~K (without SOI) and $\sim 623$~K (with SOI). Our results are consistent with those obtained using {\it ab-initio} MD predicting that the ``band convergence'' occurs at $\sim$700 K~\cite{Gibbs2013}, while an earlier MD calculation obtained the ``convergence'' temperature of $\sim 400$~K.

Fig.~\ref{sigma} also shows the individual effects of thermal expansion and electron-phonon interaction on the energy difference $\Delta$ between the VBM at L and $\Sigma$ computed using the LDA without SOI. The individual contributions to the temperature coefficient of $\Delta$ from thermal expansion and electron-phonon interaction are also given in Table \ref{slopes}. The contributions of both thermal expansion and EPI to $\frac{d\Delta}{dT}$ are negative i.e. $\Delta$ decreases with temperature. The EPI contribution to $\frac{d \Delta}{d T}$ is stronger than the thermal expansion contribution in the entire temperature range, similarly as for the direct gap. 

\begin{figure}[h]
  \begin{center}
  \includegraphics[width=8.6cm]{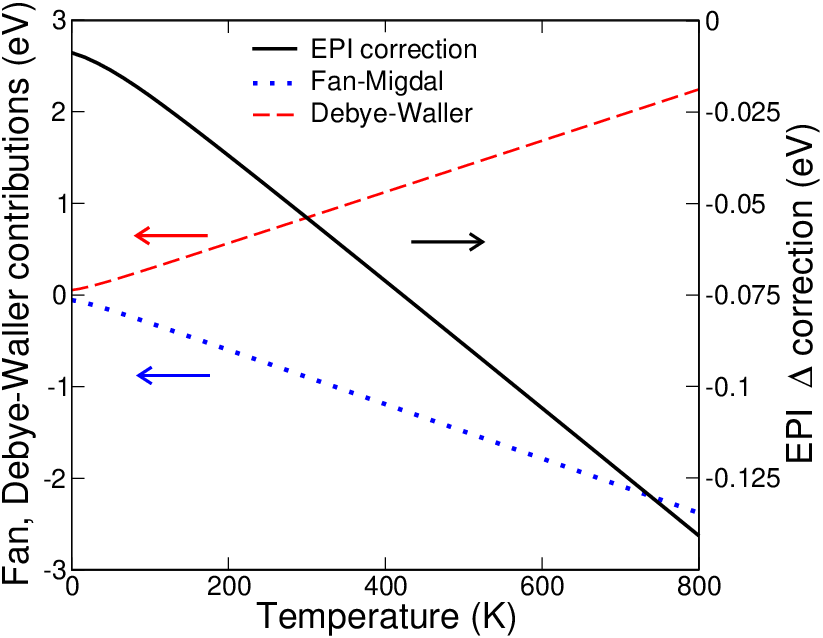}
  \caption[]{Temperature-dependent correction of the energy difference between the valence band maxima at L and $\Sigma$ in PbTe due to electron-phonon interaction (solid black line), and the Fan-Migdal (dotted blue line) and the Debye-Waller (dashed red line) contributions to electron-phonon interaction. These results are obtained using the local density approximation excluding spin-orbit interaction.}
  \label{converg}
  \end{center}
\end{figure}

\begin{figure}[h]
  \begin{center}
  \includegraphics[width=8.6cm]{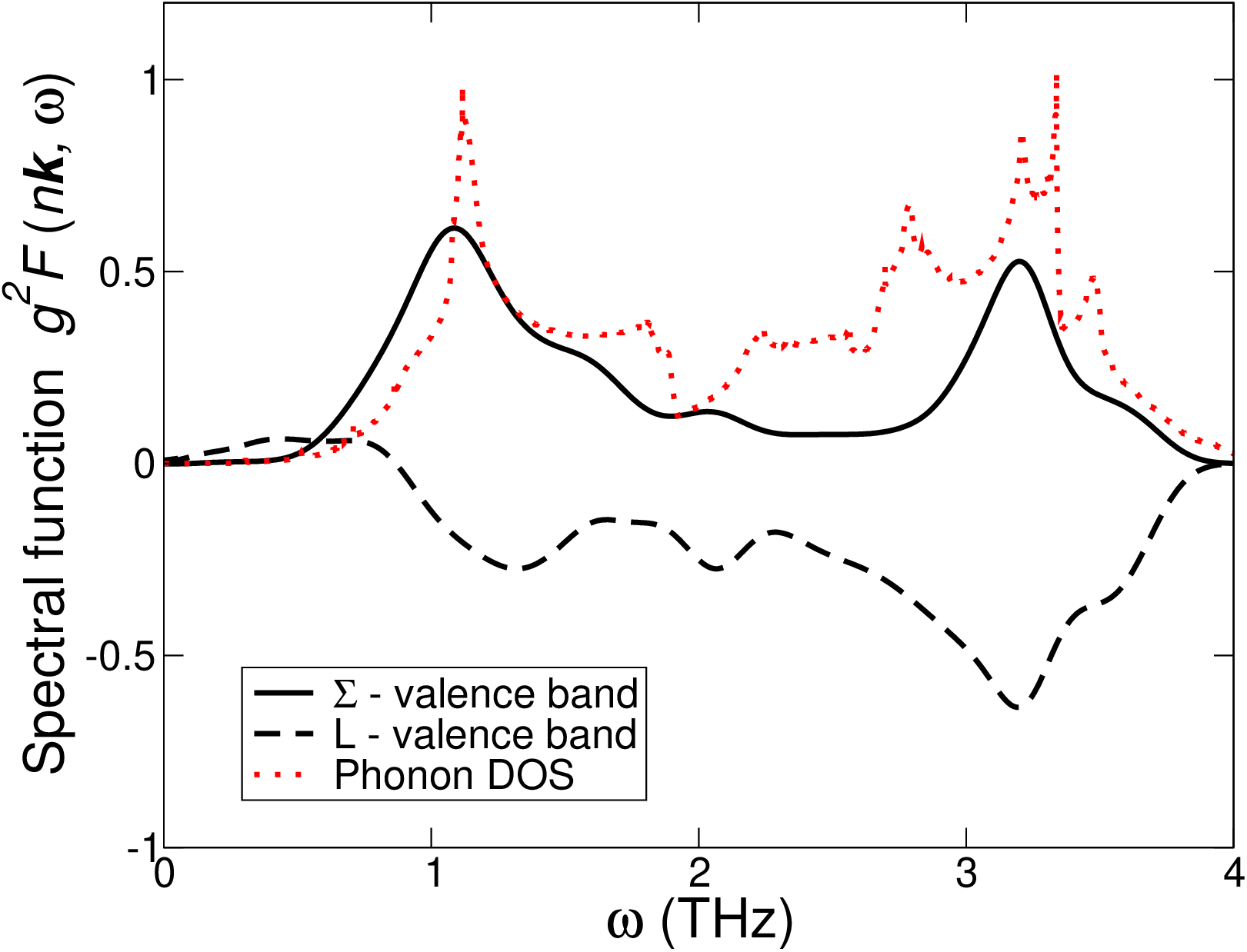}
  \caption[]{Spectral function $g^{2}F(n\one,\omega)$ (see text for explanation) versus phonon frequency for the valence band maximum at $\Sigma$ (solid black line) and L (dashed black line) in PbTe. The phonon density of states is also given by dotted red line. These results are computed using the local density approximation excluding spin-orbit interaction.}
  \label{spectralsigma}
  \end{center}
\end{figure}

We next analyse the effect of the Fan-Migdal and the Debye-Waller contributions to EPI on the temperature dependence of $\Delta$, see Fig.~\ref{converg}. The Fan-Migdal term decreases $\Delta$ with increasing temperature. The Debye-Waller term has a similar magnitude as the Fan-Migdal term, but it is smaller and has the opposite sign. The importance of including the Fan-Migdal term in determining the sign of $\frac{d \Delta}{d T}$ was also deduced in the early theoretical work of Ref.~\onlinecite{cohen1975}. Consequently, in contrast to the direct gap, the Fan-Migdal contribution to EPI and thermal expansion produce the negative sign for the temperature variation of the energy difference between the VBM at L and $\Sigma$. 

Finally, we identify the dominant phonons that contribute to the electron-phonon renormalization of $\Delta$. Fig.~\ref{spectralsigma} shows the spectral function $g^{2}F(n\one, \omega)$ for the VBM at L and $\Sigma$. High-frequency optical phonons above 3~THz give a large contribution to the electron-phonon renormalization of the VBM at L and $\Sigma$, similarly as for the direct gap. However, acoustic phonons also contribute considerably to the EPI renormalization for the VBM at $\Sigma$, in contrast to the EPI renormalization of the direct gap.

\section{Conclusions and summary}\label{conclusions}

We have investigated the temperature variation of the direct band gap and the energy difference between the L and $\Sigma$ valence band maxima of PbTe from first principles. We have analyzed the effect of electron-phonon interaction on the electronic structure renormalization using the non-adiabatic Allen-Heine-Cardona formalism and density functional perturbation theory, as well as the renormalization induced by thermal expansion using density functional theory. We obtain the temperature dependence of the direct gap of PbTe that is in very good agreement with that observed experimentally. We predict that the valence band maxima at L and $\Sigma$ become aligned at $\sim 600 - 700$ K. These parameters may be useful for building accurate models of the electronic bands and thermoelectric transport properties of PbTe. We find that both thermal expansion and electron-phonon interaction have a substantial influence on these temperature variations. Thermal expansion and the Debye-Waller (Fan-Migdal) contribution to electron-phonon interaction determine the sign of the temperature changes of the direct gap (the energy difference between the L and $\Sigma$ valence band maxima). High-frequency optical phonons are mostly responsible for the electron-phonon induced renormalization of the direct gap, whereas acoustic phonons also contribute to the ``convergence'' of the valence band maxima at L and $\Sigma$.

\section{Acknowledgements}

We thank Felipe Murphy-Armando, Djordje Dangi\'c, Aoife R. Murphy, and Tchavdar Todorov for helpful discussions. J.~D. Q.-F. acknowledges Samuel Ponc\'e for helpful suggestions on technical aspects of the {\sc ABINIT} code. This work was supported by Science Foundation Ireland under Investigators Programme No. 15/IA/3160. We acknowledge the use of computational facilities at the Irish Centre for High-End Computing (ICHEC).
 
\appendix
\section{Linear thermal expansion coefficient, lattice constant and phonon dispersion of PbTe}\label{newappA}

\begin{figure}[h]
  \begin{center}
  \includegraphics[width=8.6cm]{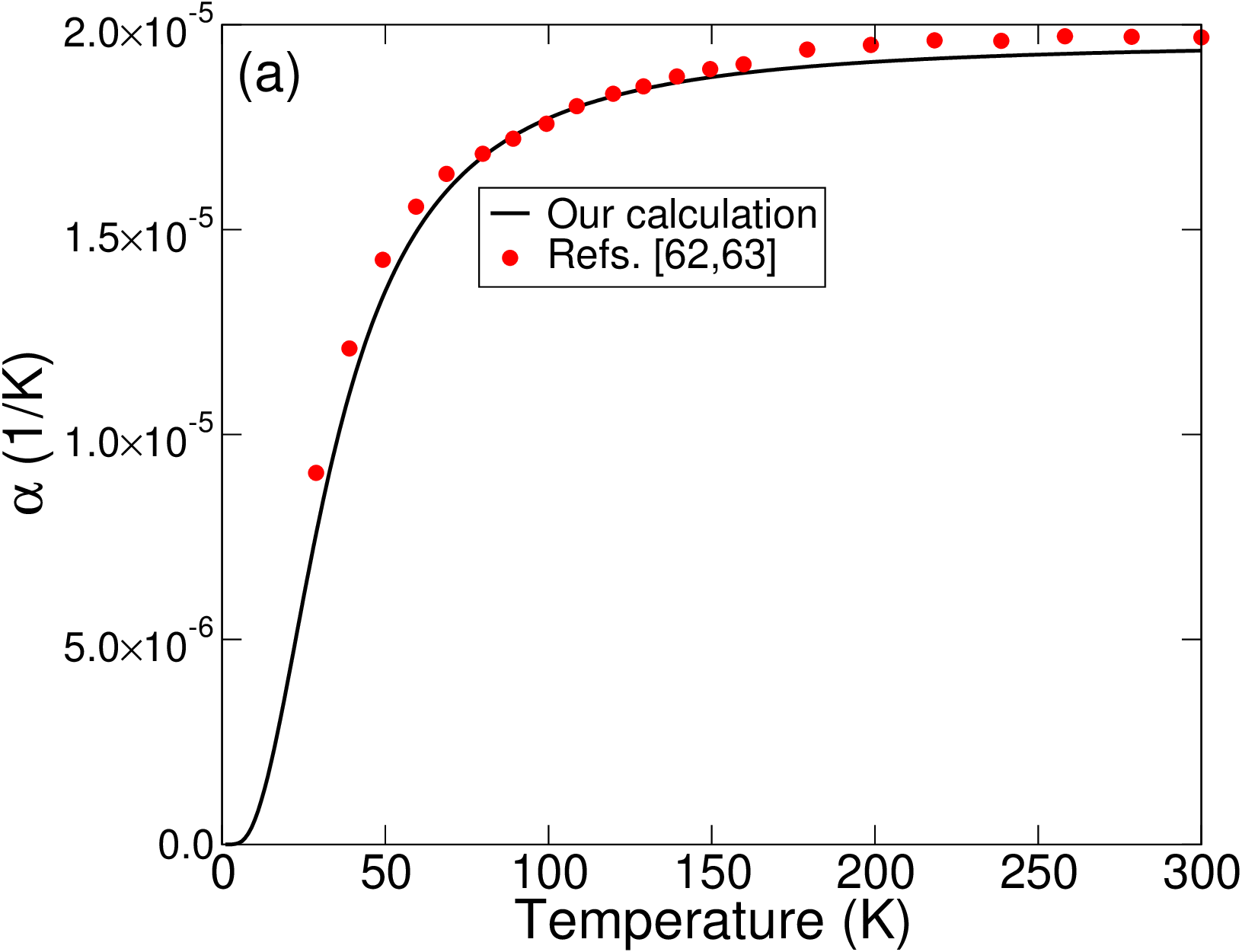}\\
  \includegraphics[width=8.1cm]{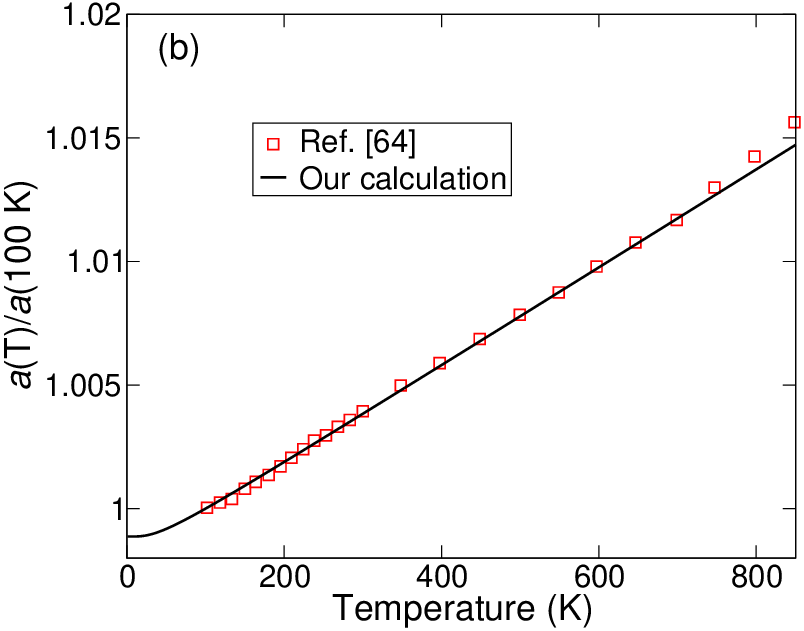}
  \caption[]{(a) Calculated (solid black line) and experimental~\cite{novikova,cochran1966} (red dots) linear thermal expansion of PbTe. (b) Calculated (solid black line) and experimental~\cite{Kastbjerg} (red squares) temperature-dependent lattice constant of PbTe divided by its value at $100$~K. The local density approximation without spin-orbit interaction is used in these calculations.}
  \label{lattice}
  \end{center}
  \end{figure}
  
\begin{figure}[h]

  \begin{center}
  \includegraphics[width=8.2cm]{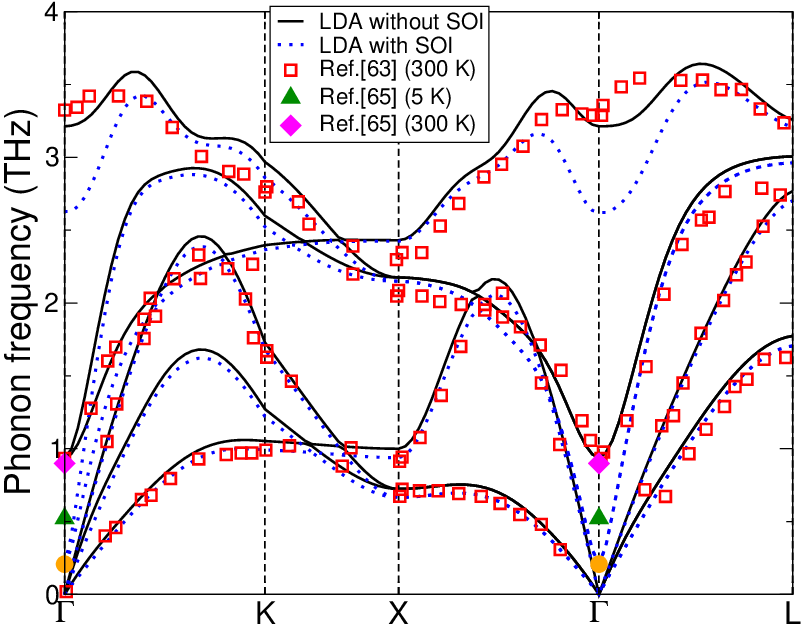}
  \caption[]{Phonon dispersion of PbTe calculated using density functional perturbation theory and the local density approximation excluding (solid black line) and including (dotted blue line) spin-orbit interaction. Orange circles indicate the frequency of the zone center transverse optical mode when spin-orbit interaction is included. Experimental data from inelastic neutron scattering~\cite{cochran1966} (red squares) and optical spectroscopy~\cite{bauer} (magenta diamonds and green triangles) are also shown.}
  \label{phonons}
  \end{center}
  \end{figure}  
  
 \begin{figure}[h]
  \begin{center}
  \includegraphics[width=8.6cm]{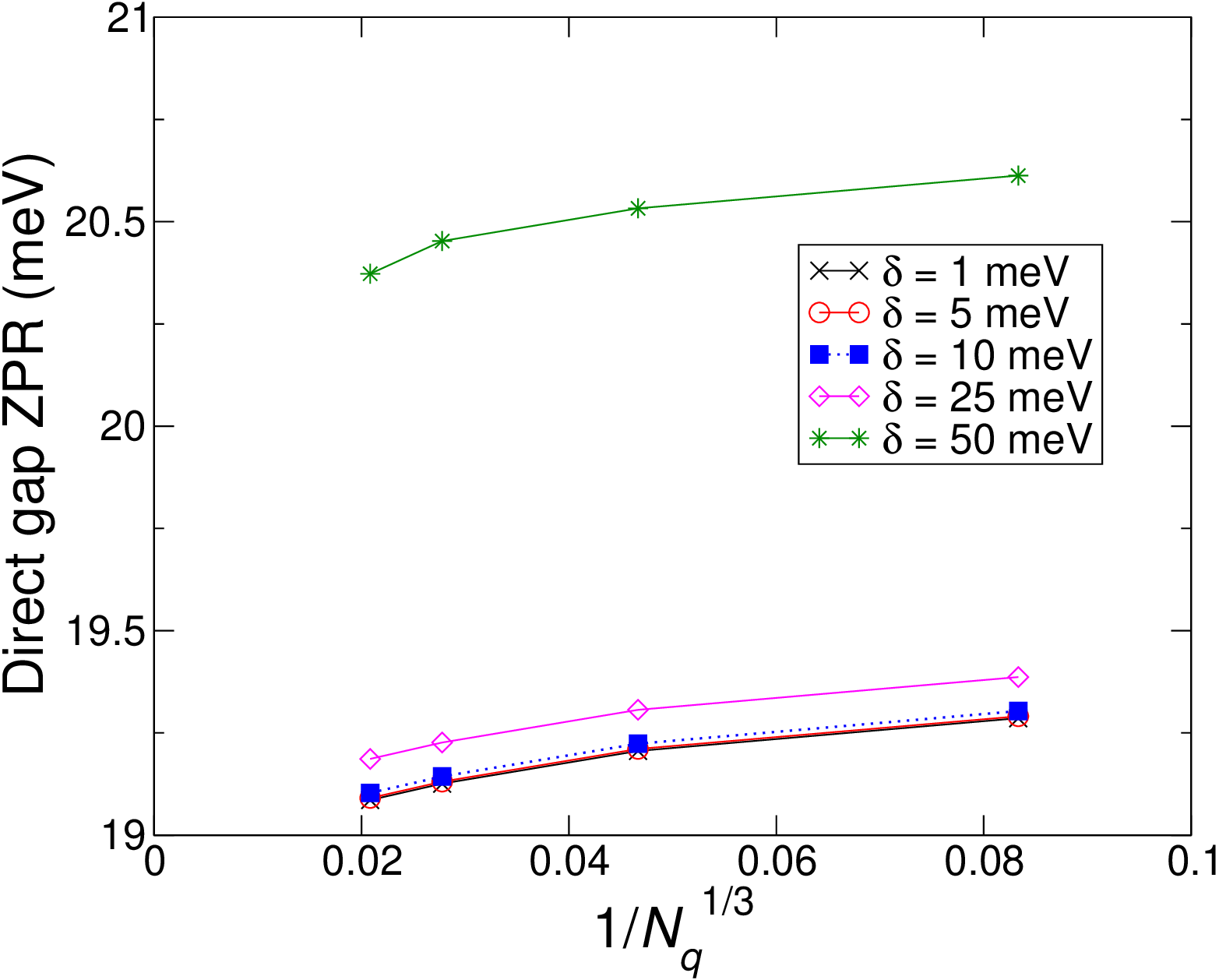}
  \caption[]{Convergence study for the zero-point renormalization (ZPR) of the direct band gap of PbTe with respect to the $\two$-point grid density  ($N_{\two}$ is the total number of sampled $\two$-points). These results are computed using the local density approximation excluding spin-orbit interaction.}
  \label{ZPRdirect}
  \end{center}
  \end{figure}

\begin{figure*}[ht]
\begin{center}
\includegraphics[width=0.31\linewidth]{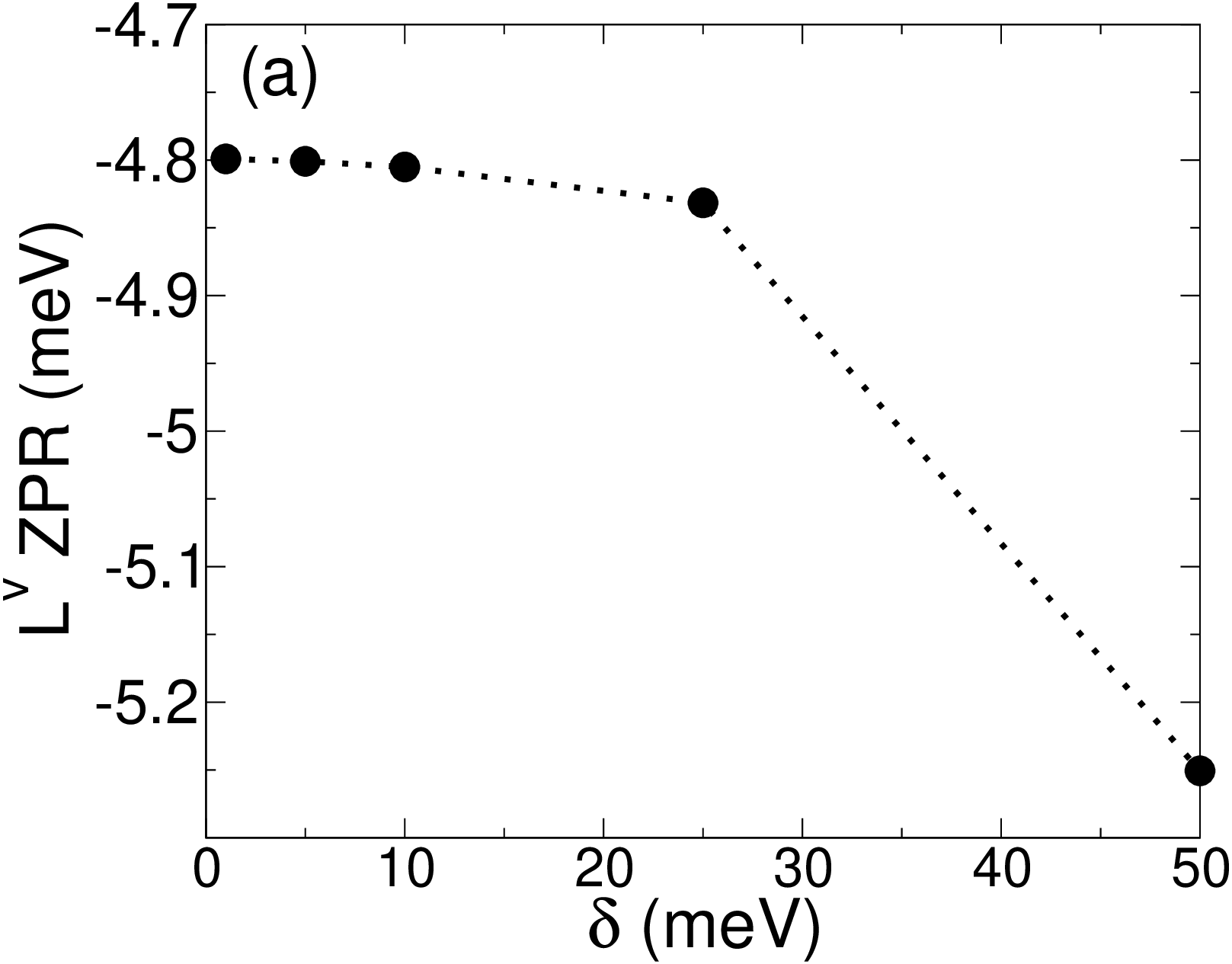} \hspace{3mm}
\includegraphics[width=0.315\linewidth]{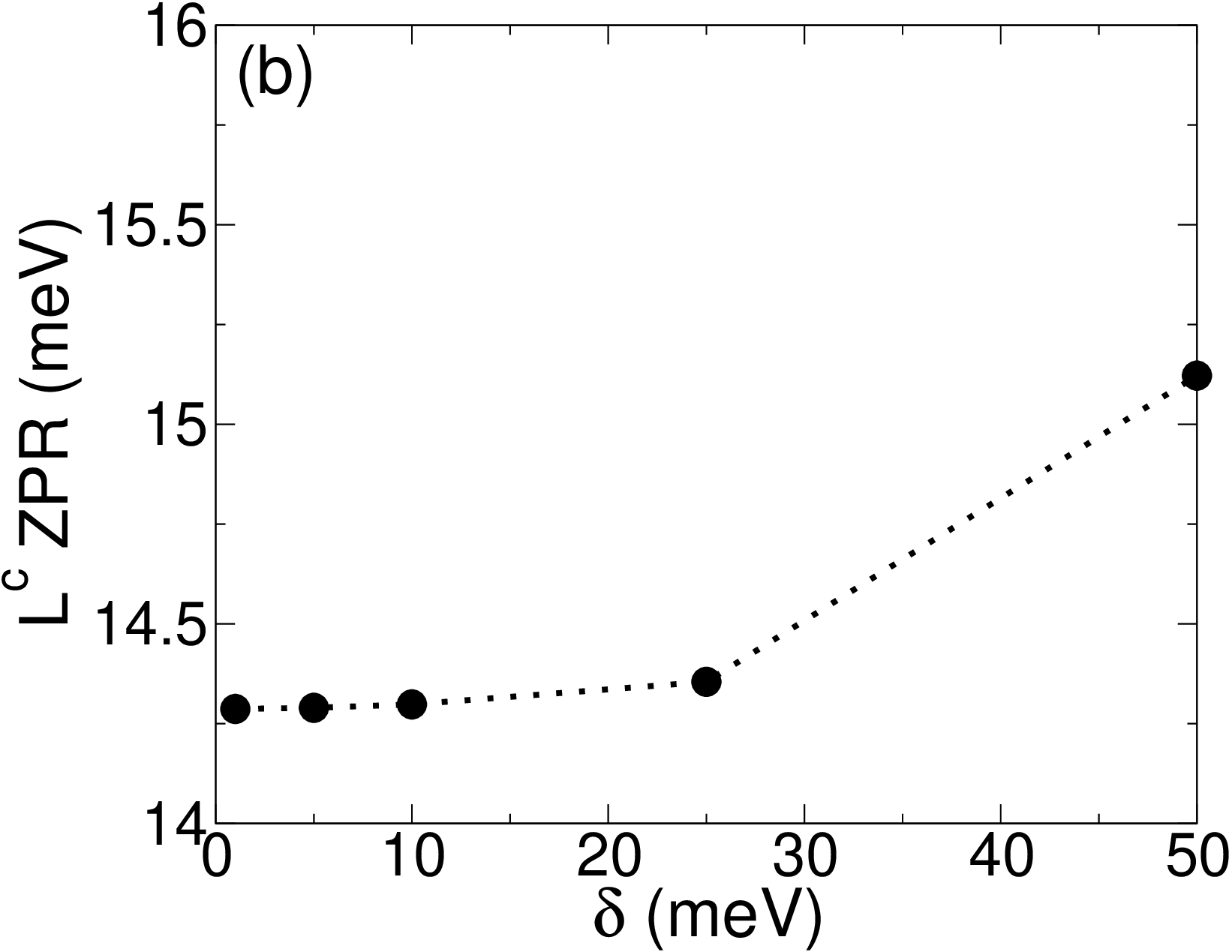} \hspace{3mm}
\includegraphics[width=0.31\linewidth]{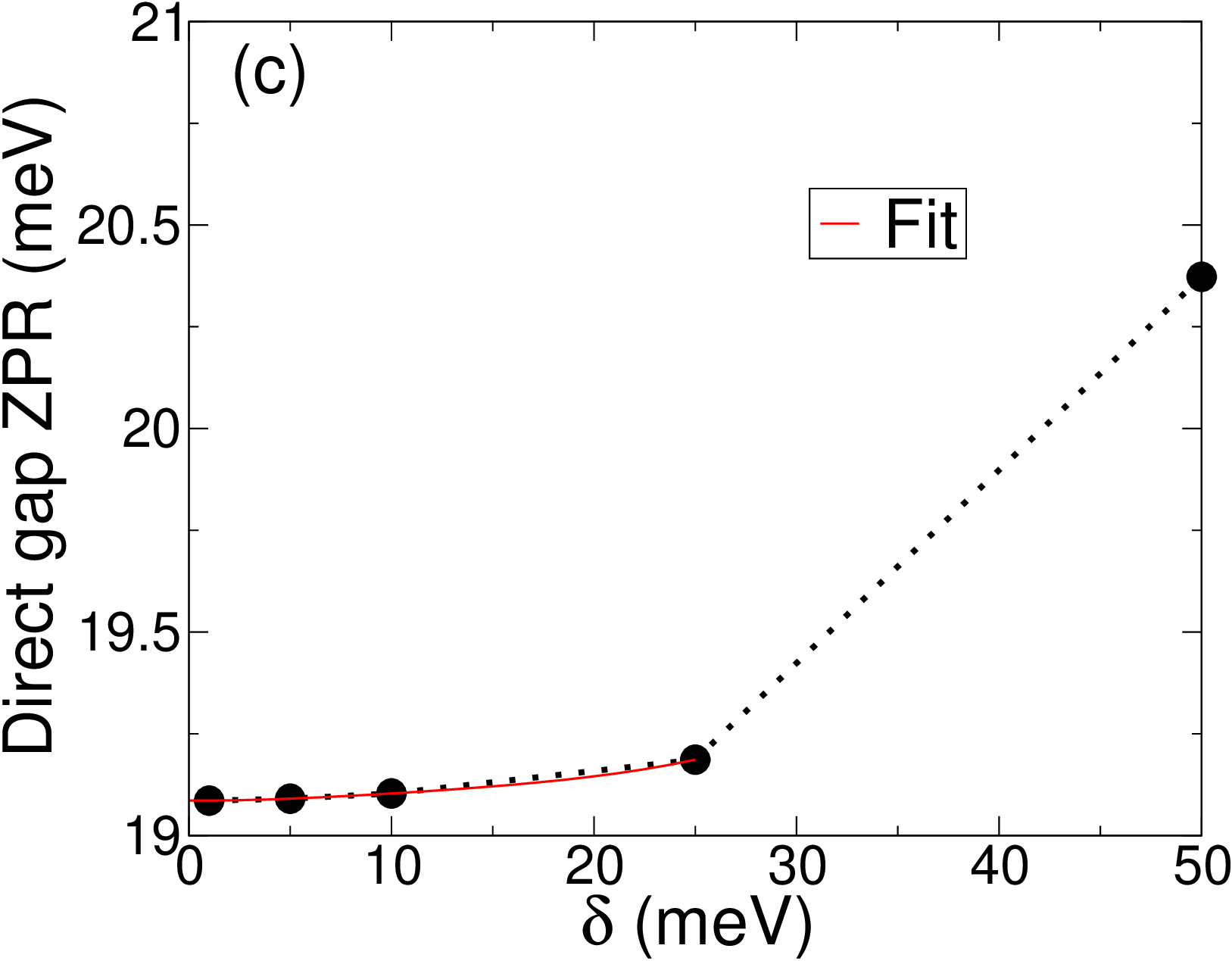}  
\end{center}
\caption{Convergence study for the zero-point renormalization (ZPR) at the L point of: (a) the valence band (L$^{v}$), (b) the conduction band (L$^{c}$) and (c) the direct gap ($E_g$) with respect to the 
parameter $\delta$ for the $\two$-point grid of 48 $\times$ 48 $\times$ 48. The Lorentzian fit for $\delta\rightarrow0$ for the ZPR of the direct gap is shown by solid red line. The local density approximation without spin-orbit interaction is used in these calculations.}
\label{delta}
\end{figure*}   

Fig. \ref{lattice} shows our calculated linear thermal expansion and the lattice constant of PbTe as a function of temperature using the LDA without SOI. The computed values agree very well with experiments~\cite{novikova,cochran1966,Kastbjerg}. The LDA lattice constant is 6.348 \AA. When the zero point renormalization is taken into account, we obtain the lattice constant of 6.355 \AA ~at 0 K. Our calculated lattice constant at 300 K is 6.386 \AA, which compares fairly well with the experimental values of 6.46179 \AA \cite{Kastbjerg} and 6.462 \AA \cite{dalven}. The lattice constant obtained using the LDA including SOI does not differ much from the one without SOI (6.339 \AA, $\sim$0.14$\%$ smaller than the one without SOI).
  
In Fig. \ref{phonons}, we plot the phonon dispersion of PbTe calculated using DFPT and LDA, and compare it with the experimental data from inelastic neutron scattering (INS) at 297 K \cite{cochran1966} and optical spectroscopy at 5 K and 300 K~\cite{bauer}. The phonon band structure calculated without SOI agrees very well with that measured with INS. The TO mode frequency is closer to the optical measurements at 5 K when SOI is accounted for. This softening of the TO mode due to SOI was also observed in the previous DFPT-LDA calculations of Ref.~\onlinecite{romero2008}, and can be explained by the gap inversion and a resulting strong modification of the electron-phonon coupling between valence and conduction bands~\cite{nature}. Other than the TO mode close to zone center, the phonon band structures calculated using the LDA with and without SOI are very similar.

\section{Convergence study for the zero-point renormalization of the direct gap due to electron-phonon interaction}\label{adiabatic} 

Non-adiabiatic effects on the temperature dependence of the electronic band structure can be accounted for by keeping the phonon frequencies $\omega_{\two \lambda}$ in the Fan-Migdal self-energy given by Eq.~\eqref{selfenergy}. The convergence study for the ZPR of the direct gap at L with respect to the $\two$-point grid density calculated using the non-adiabatic AHC approach is given in Fig.~\ref{ZPRdirect}. The direct gap converges linearly with 1/$N_{\two}^{1/3}$, where $N_{\two}$ is the total number of sampled $\two$-points. The results appear nearly converged for the $\two$-grid densities of 36 $\times$ 36 $\times$ 36 (1/$N_{\two}^{1/3}$ = 0.0278) and 48 $\times$ 48 $\times$ 48 (1/$N_{\two}^{1/3}$ = 0.0208), and small $\delta$ values ($\delta\le 10$~meV). Nevertheless, the long-range ($\two\rightarrow 0$) longitudinal optical (LO) phonon contribution to the band structure renormalization is not fully accounted for in these calculations (see Appendix~\ref{polaron}). In Fig. \ref{delta},  we show the convergence study with respect to $\delta$ for the valence band ($L^{v}$), conduction band ($L^{c}$) and direct gap at L and the $\two$-grid of 48 $\times$ 48 $\times$ 48. Using a Lorentzian fit for $\delta\rightarrow0$ as discussed in Ref.~\onlinecite{ponce2015jcp}, we calculate the direct gap ZPR of 19.09 meV using the LDA without SOI. We have checked this result including SOI with the $\two$-grid of 48 $\times$ 48 $\times$ 48 and $\delta =1$~meV, and we obtain the ZPR value for the direct gap of -21.58 meV.

\section{Polaronic contribution to the band structure renormalization}\label{polaron} 

It has been pointed out that the densities of commonly used $\two$-grids in non-adiabatic AHC-DFPT calculations (of the order of 48$\times$48$\times$48) may not be sufficiently large to accurately describe the long-range LO phonon (polaronic) contribution to the band structure renormalization as $\two\rightarrow 0$ \cite{nery2016}. We estimate this error in our calculations on a 48$\times$48$\times$48 $\two$-grid in the following manner: we first calculate the polaronic shift on very dense $\two$-grids using the effective mass approximation and the Fr\"ohlich model for electron-LO phonon coupling (similarly to Ref. \onlinecite{nery2016}). We then subtract the corresponding polaronic shift obtained using the 48$\times$48$\times$48 $\two$-grid we used in the DFPT calculation from the converged shift on a very dense grid. 

The polaronic shift expression we used in the calculations above is a generalization of Eq.~(2) in Ref.~\onlinecite{nery2016} for the cases where the relevant electronic band states are described by two or three effective masses (CBM and VBM of PbTe at L, and VBM at $\Sigma$, respectively). For example, the polaronic shift of the conduction band at L can be given as:
\begin{eqnarray}\label{polaronic_shift}
  \Delta E_{\text{L}}^{\text{CBM}} = - \frac{1}{N_{\two}} \sum_{\two} \frac{\hbar e^2 \omega_{\rm LO}}{2 V \epsilon_0} \left( \frac{1}{\epsilon_\infty} - \frac{1}{\epsilon_s} \right)\frac{1}{q^2}  \times  \nonumber\\ \left[ \frac{n_{\rm LO}(T)+1}{\hbar^2 q_\parallel^2/2m_\parallel^* + \hbar^2 q_\perp^2/2m_\perp^* + \hbar \omega_{\rm LO}} + \right.\nonumber\\ \left.\frac{n_{\rm LO}(T)}{\hbar^2 q_\parallel^2/2m_\parallel^* + \hbar^2 q_\perp^2/2m_\perp^* - \hbar \omega_{\rm LO} } \right],
\end{eqnarray}
where $\omega_{\rm LO}$ is the LO phonon frequency (approximately taken as a constant for different \two), $\epsilon_0$ the vacuum permittivity, $\epsilon_\infty$ and $\epsilon_s$ the high-frequency and static dielectric constant, $\hbar$ the reduced Planck constant, $e$ the electron charge, and $V$ the unit cell volume. $n_{\rm LO}(T)$ is the Bose-Einstein occupation for the LO phonons. $q_\parallel$ and $q_\perp$ are the projections of the $\two$ vector on the L-$\Gamma$ and L-W directions, which are parallel to the directions of parallel and perpendicular effective masses at L, $m_\parallel^*$ and $m_\perp^*$. All these parameters were obtained from our DFT and DFPT calculations \cite{ronan2018,jiang2018}. Our calculated effective masses of the $\Sigma$ valley in the units of free-electron mass along the three principal axes $[110]$, $[1\bar{1}0]$, and $[001]$ are: $m_{\parallel}=0.178$,   $m_{\perp xy}=0.046$ and  $m_{\perp z}=3.788$, and agree well with the corresponding QSGW values~\cite{GW}.

\begin{figure}[h]
\begin{center}
\includegraphics[width=8.2cm]{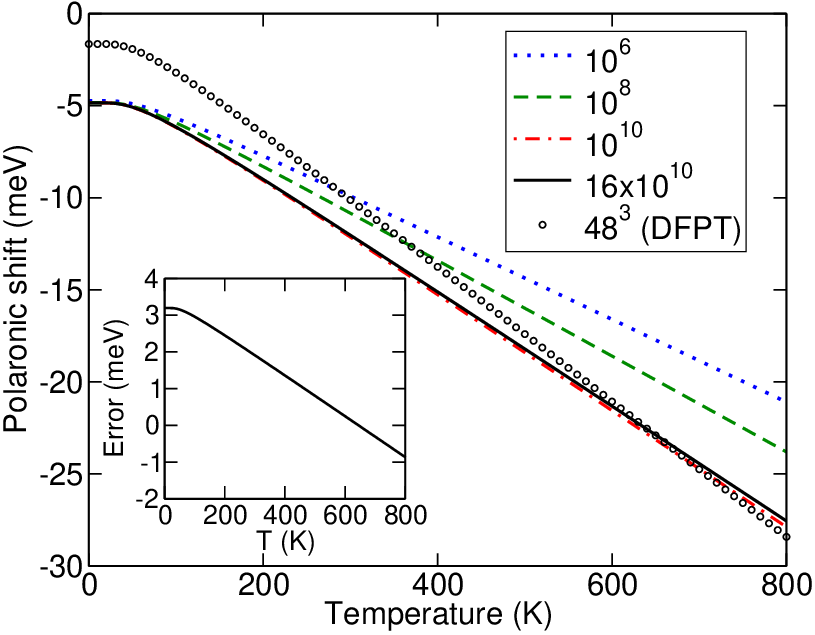} \hspace{3mm}
\end{center}
\caption{Convergence of the polaronic shift for the direct gap at L up to 800~K with respect to the number of $\two$ points. Black circles represent the polaronic shift calculated on a 48$\times$48$\times$48 $\two$-grid used in our density functional perturbation theory (DFPT) calculations. Inset: Energy difference between the converged polaronic shift of the direct gap (solid black line in the main figure) and the one obtained using the 48$\times$48$\times$48 DFPT \two-grid.}
\label{error}
\end{figure}

The second term of Eq.~\eqref{polaronic_shift} becomes divergent as the \two-grid density increases. The singularity in this term is computed by principal parts integration. To do that, we center the \two-grids with respect to the pole of the integrand. Fig.~\ref{error} shows the convergence of the polaronic shift for the direct gap at L of PbTe with respect to the \two-grid density. We find that very dense \two-grids with $\sim 10^{10}$ points are needed to converge the direct gap values. Similarly, we find that it is necessary to use $\sim 10^{12}$ \two-points to converge the energy difference $\Delta$ between the VBM at L and $\Sigma$ for PbTe. We note that the principal part approach on a fine \two-grid presented here might not be the only way to converge the polaronic contribution to energy shifts. An analytic solution of Eq.~\eqref{polaronic_shift} may allow an accurate answer with a coarser \two-grid, as done in Ref.~\onlinecite{nery2016} for the isotropic effective mass case.

In addition to the converged polaronic shift for the direct gap at L, Fig.~\ref{error} also shows the corresponding shift for the 48x48x48 DFPT $\two$-grid, and their difference in the inset, which gives our estimated errors. These errors for the direct gap and those for the energy difference  between the VBM at L and $\Sigma$ are relatively small compared to the total shifts calculated on the 48x48x48 DFPT $\two$-grid and given in Figs.~\ref{totaldirectgap} and \ref{sigma}, respectively. For the direct gap at L, the errors range from 12.9\% at 0 K to 0.32\% at 800 K. For the energy difference  between the VBM at L and $\Sigma$, they range from 10.74\% at 0 K to 1.93\% at 800 K. 

\bibliography{references} 

\end{document}